\documentclass[
    ,final            
  ]
  {aipproc}

\layoutstyle{6x9}

\begin{document}

\title{Ascertaining the Core Collapse Supernova Mechanism: An Emerging Picture?}


\author{Anthony Mezzacappa}{
  address={Physics Division, Oak Ridge National Laboratory, Oak Ridge, TN 37831}
}

\author{Stephen W. Bruenn}{
  address={Department of Physics, Florida Atlantic University, Boca Raton, FL 33431}
}

\author{John M. Blondin}{
  address={Department of Physics, North Carolina State University, Raleigh, NC 27695}
}

\author{W. Raphael Hix}{
  address={Physics Division, Oak Ridge National Laboratory, Oak Ridge, TN 37831}
}

\author{O.E. Bronson Messer}{
  address={National Center for Computational Sciences, Oak Ridge National Laboratory, Oak Ridge, TN 37831}
}

\begin{abstract}
The mechanism for core collapse supernova explosions remains undefined in detail and perhaps even in broad brush. Past multidimensional simulations point to the important role neutrino transport, fluid instabilities, rotation, and magnetic fields play, or may play, in generating core collapse supernova explosions, but the fundamental question as to whether or not these events are powered by neutrinos with the aid of some or all of these other phenomena or by magnetic fields or by a combination of both is unanswered. Here we present the results from two sets of simulations, in two and three spatial dimensions. In two dimensions, the simulations include multifrequency flux-limited diffusion neutrino transport in the ``ray-by-ray--plus'' approximation, two-dimensional self gravity in the Newtonian limit, and nuclear burning through a 14-isotope alpha network. The three-dimensional simulations are model simulations constructed to reflect the post stellar core bounce conditions during neutrino shock reheating at the onset of explosion. They are hydrodynamics-only models that focus on critical aspects of the shock stability and dynamics and their impact on the supernova mechanism and explosion. The two-dimensional simulations demonstrate the important role nuclear burning may play despite the relatively small total energy deposition behind the shock. The three-dimensional simulations demonstrate the need for three-dimensional multi-physics core collapse supernova models. In two dimensions, with the inclusion of nuclear burning, we obtain explosions (although in one case weak) for two progenitors (11 and 15 M$_{\odot}$ models). Moreover, in both cases the explosion is initiated when the inner edge of the oxygen layer accretes through the shock. Thus, the shock is not revived while in the iron core, as previously discussed in the literature. The three-dimensional studies of the development of the stationary accretion shock instability (SASI) demonstrate the fundamentally new dynamics allowed when  simulations are performed in three spatial dimensions. The predominant $l=1$ SASI mode gives way to a stable $m=1$ mode, which in turn has significant ramifications for the distribution of angular momentum in the region between the shock and proto-neutron star and, ultimately, for the spin of the remnant neutron star. Moreover, the three-dimensional simulations make clear, given the increased number of degrees of freedom, that two-dimensional models are severely limited by artificially imposed symmetries.
\end{abstract}

\maketitle

\section{Introduction}

Spherically symmetric models of core collapse supernovae have achieved a significant level of sophistication and realism. For example, general relativistic models with Boltzmann neutrino transport and sophisticated weak interaction physics have now been performed \cite{lmtmhb01}. While explosions are not obtained, these simulations nonetheless laid the foundation for later two- and three-dimensional simulations, continue to provide a reference point for the interpretation of these later multidimensional models, and in and of themselves contributed in a significant way to core collapse supernova theory. That explosion is not obtained in spherical symmetry demonstrates that multidimensional effects are required for explosion, which is an important conclusion that frames all subsequent work in two and three dimensions differently than if the opposite were true. Similarly, two-dimensional models have now also achieved a significant level of realism and sophistication. These models include multifrequency neutrino transport via flux-limited diffusion or variable Eddington factor approximations in the ``ray-by-ray--plus'' approximation \cite{burasetal06,bruennetal06} or two-dimensional multigroup flux-limited diffusion
\cite{sm05b,burrowsetal06}. Unlike in the spherically symmetric models, the outcomes in the two-dimensional models vary qualitatively and quantitatively, which reflects the increased complexity present in two-dimensional models, the increase in the number of components involved in modeling core collapse supernovae in two dimensions and the range of approximations that have been employed in modeling these components, and the increased need for model precision in a tightly coupled nonlinear multidimensional, rather than one-dimensional, problem. In this regard, there is one comparison of note: In two-dimensional models with ray-by-ray--plus neutrino transport, general relativistic corrections to the gravitational field, and nuclear ``flashing,'' explosions are not obtained by one group except for an 11 M$_{\odot}$ progenitor \cite{burasetal06}, and in this case only a very weak explosion is obtained. On the other hand, in comparable two-dimensional models by our group \cite{bruennetal06} and presented here, with nuclear burning described by an alpha network of nuclei but Newtonian gravity, explosions are obtained for progenitors of both 11 and 15 M$_{\odot}$, although here too, as we will discuss, the explosion in the latter case is weak. Simulations are now underway with general relativistic corrections to the gravitational field. The qualitative and quantitative differences in these two sets of simulations by two different groups may arise from the increased sophistication in the way nuclear burning is treated or the decreased sophistication with which the gravitational field is treated in one or the other of the two approaches. The future simulations mentioned above will be reported elsewhere and will be able to determine whether the increased sophistication in the modeling of the nuclear burning remains a decisive factor for the approximate general relativistic runs. Here we will show that, in the Newtonian limit, this is certainly the case. 

Nonetheless, the consistent lack of explosions in the spherically symmetric case, despite the sophistication and the realism of the models (within the framework of an unrealistic imposed symmetry, of course), and the now mixed outcomes being obtained in ever more sophisticated two-dimensional models is heartening. There is the hope that three-dimensional models with all the known phenomena mentioned above, each treated with sufficient realism, will unlock the details of the process through which massive stars die. Of course, it is also possible, but perhaps not likely, we are still missing a major piece of the physics. In either event, there is much work to be done to include all of the known relevant physics in three dimensions with sufficient accuracy. At that point, depending on the outcome, we would be in a better position to answer such final questions.

\section{Results}

Here we report on two sets of simulations: one in two dimensions, the other in three. In the first of these, two-dimensional simulations were performed with two-dimensional hydrodynamics coupled to ``ray-by-ray-plus'' \citep[cf.][]{buras_rjk06} multigroup flux-limited diffusion neutrino transport and a nuclear  
reaction network \cite{HiTh99b} of 14 alpha nuclei between helium and zinc. The equation of state (EOS) of Lattimer and Swesty \citep{LS91} was used for matter in NSE above $1.7 \times 10^{8}$ g cm$^{-3}$. Below this density, matter in NSE is described by 4 elements (neutrons, protons, helium, and a representative heavy nucleus) in a modified version of the EOS described by \citep{cooperstein85a}. 
For regions not in NSE, the nuclei are treated as an ideal gas. 
The ideal gas consists of 14 alpha nuclei ($^{4}$He to $^{60}$Zn) evolved by a nuclear  
reaction network, and three additional species: n, p, and neutron-rich 
iron ($^{56}$Fe), which are included to affect a smooth freeze-out of 
matter from NSE.
An electron--positron EOS with arbitrary degeneracy and degree of relativity spans the entire density--temperature range in our simulations. The neutrino opacities we use are the ``standard'' opacities described in \citep{br85}, with the isoenergetic scattering of nucleons replaced by the more exact formalism of \citep{rpl98}, which includes nucleon blocking, recoil, and relativistic effects, and with the addition of nucleon--nucleon bremsstrahlung \citep{hr98}, with the kernel reduced by a factor of five in accordance with the results of \citep{hpr01}. Details on our numerical methods and the code can be found elsewhere \cite{bruennetal06}. 

Our two-dimensional simulations were initiated from the 11 M$_\odot$ and 15 M$_\odot$ progenitors S11s7b and S15s7b provided by Woosely \cite{w06}. They were carried out to $\sim 700$ ms after bounce. As had been pointed out elsewhere for different reasons \cite{burrowsetal06}, pushing the simulations beyond the first few hundred milliseconds after bounce is crucial in order to capture all of the post-stellar core bounce dynamics---particularly, the correct final outcome. 

For both stellar progenitors, the following picture emerges: After bounce, the shock stalls, and there is an initial period of Ledoux convection driven by  the negative entropy gradient left behind by the weakening shock. At $\sim$ 50 ms after bounce, this convective phase is followed by a phase of neutrino-driven convection between the gain radius and the shock, driven by the newly established entropy gradient there due to neutrino heating. Neutrino-driven convection is fully established by $\sim$ 100 ms after bounce. Thus far we have described the standard scenario detailed by numerous authors in the past \cite{hbhfc94,bhf95,jm96,mcbbgsu98b,burasetal103,fw04,burasetal06}. By 140 (200) ms after bounce in the 11 (15) M$_\odot$ case, the shock begins to exhibit quasi-oscillatory global distortions along the polar axis, characteristic of the $l=1$ SASI mode. Then, when the inner boundary of the oxygen layer reaches the shock, a fundamental change in the dynamics occurs. Due to the global shock distortion from the SASI, this first occurs along the polar axis. At this stage, an explosion is initiated for both progenitors, although the delay to explosion is different in each case, dictated largely by the very different radial locations of the inner boundary of the oxygen layer and the time it takes for this layer to reach the shock. For the 11 M$_\odot$ progenitor, the 0.1 and 0.5 mass fractions of oxygen are at 1400 km and 1600 km, respectively. For the 15 M$_\odot$ case, they are instead at 2500 km and 6400 km, respectively. The delay to explosion is $\sim 160$ ms for the 11 M$_\odot$ model and $\sim$450 ms for the 15 M$_\odot$ model. 

Of particular note is the fact that, for both models, if the nuclear network is turned off an explosion does not occur. The impact of nuclear burning is not subtle in the sense that it alters qualitatively the outcome in the simulations, but it is subtle in the sense that the total energy released through nuclear burning is small ($< 0.1$ B). When the network is turned off, instantaneous conversion (``flashing'') of oxygen or silicon to iron in a computational zone occurs when the {\it ad hoc} $T>0.44$ MeV criterion for the transition to NSE is met, which may or may not occur close to the shock (ahead of or behind it), where we expect burning actually to occur. On the other hand, when an alpha network is used, burning will occur in the vicinity of the shock, and in this case, partial burning with the release of some nuclear energy is also possible, which may be particularly important for weak shocks. It is this more effective deposition of nuclear energy that we believe qualitatively alters the outcome in our models. This increases the postshock pressure and causes the shock to move farther out. This, in turn, increases the important ratio of the advective time scale to the neutrino heating time scale \citep[see][]{burasetal06} to values exceeding unity, and the explosion is initiated. Thus, it is the confluence of neutrino heating, the SASI, effective nuclear energy deposition in the vicinity of the shock, and the decreased ram pressure in the oxygen layer that leads to explosions in our models when an alpha network is deployed. (Note that flashing still occurs at the transition to NSE in models in which the nuclear network is used, but its effect is limited. When a network is used, flashing ensures that the composition predicted by the network is in agreement with the composition predicted by the EOS for material in NSE. If a compositional adjustment is needed to bring the network prediction into accord with the EOS prediction for material in NSE, some flashing occurs, but the iron-rich composition resulting from the network is much closer to NSE than the progentior composition. Hence, the nuclear energy generation (or loss) associated with this flashing is small.)

For the 11M$_\odot$ model, the explosion is nearly unipolar, with a strong inflow becoming established near the polar axis (this is evident in Figure 1). On the other hand, for the 15M$_\odot$ model the shock expansion is more bipolar, with a strong inflow established closer to the equator. A unipolar explosion will result in a larger neutron star kick, and the emergence of a unipolar versus a bipolar explosion may be stochastic and result in a bimodal pulsar velocity distribution, as pointed out by \citep{schecketal06}. The estimated explosion energies in our models at the time our simulations were terminated were $\sim$1.0 B  for the 11 M$_\odot$ case (1 B $=$ 1 Bethe $=1\times10^{51}$ erg) and $\sim$0.3 B for the 15 M$_\odot$ case. The latter is certainly weak, although the explosion energy was still increasing when the simulation was terminated.

%
In our three-dimensional models, we begin with a spherically symmetric stationary accretion shock in conditions that correspond to the postbounce stellar core at the onset of explosion. At this time, the postshock region is dominated by relativistic radiation pressure, and there is a narrow cooling layer at the base of the region between the shock and the proto-neutron star. In our simulations, these conditions are modeled with polytropic equations of state (see also \cite{janka01}) and a prescription for the cooling layer originally specified in \cite{hc91}. For more details, the reader is referred to \cite{bmd03}. Our aim in these simulations is to understand the development of the SASI in three dimensions and its impact on the shock and explosion dynamics. As in our two-dimensional studies \cite{bmd03}, the SASI begins as an $l=1$ instability. The SASI leads to the formation of an internal shock in the region below the supernova accretion shock and perpendicular to the accretion shock. (This can be seen in Figure 3 at the interface between the two oppositely directed flows.) The junction of the two shocks is a Mach reflection (the kink in the accretion shock surface in Figure 3) that in the axisymmetric case forms a ring about the polar axis on the surface of the accretion shock. When the $l=1$ mode dominates the flow, the ring moves along this surface with respect to polar angle. (This is evident in the first  two panels of Figure 2.) In two dimensions, with axisymmetry imposed, the ring converges in an axisymmetric fashion at the pole, and the reflection maintains axisymmetry as the ring begins its motion toward the opposite pole. In three dimensions, however, with the restriction of axisymmetry removed, the Mach reflection convergence is not precisely axisymmetric, nor is the reflection, and axisymmetry is broken. (This is evident in the last two panels of Figure 2.) With time, this leads to a one-sided internal shock with respect to azimuthal angle, moving clockwise if viewed from above ($\theta = 0$) on an equatorial slice, and a Mach reflection that subtends only a portion of the accretion shock surface, not a complete ring encircling the polar axis. This non-axisymmetric internal shock then has a dramatic impact on the post-accretion shock flow. The internal shock leads to two counter rotating flows, one, directly below the accretion shock, moving in a clockwise fashion with the internal shock, and another, moving in a counterclockwise fashion when viewed from above in the same equatorial plane, directly above the proto-neutron star (see Figure 3). This inner flow is formed when material infalling through the accretion shock ahead of the internal shock is subverted by the internal shock to regions above the proto-neutron star surface. The counterclockwise rotation is induced by this subversion. In turn, this inner counterclockwise rotating flow will deposit angular momentum onto the proto-neutron star as it accretes. Calculations of the resultant spin period of the remnant neutron star \cite{bm05Nature} yield a spin period $\sim 50$ ms, consistent with observations and obtained even with spherically symmetric initial conditions (the $m=1$ SASI mode is a robust outcome regardless of the initial conditions used---i.e., whether or not one begins with a rotating or nonrotating progenitor). This remarkably different angular momentum distribution below the shock relative to the two-dimensional case and its resultant impact on the supernova and remnant neutron star also demonstrates very clearly the need to extend our two-dimensional multi-physics supernova models described above to three dimensions.

\begin{figure}
\includegraphics[height=.9\textheight]{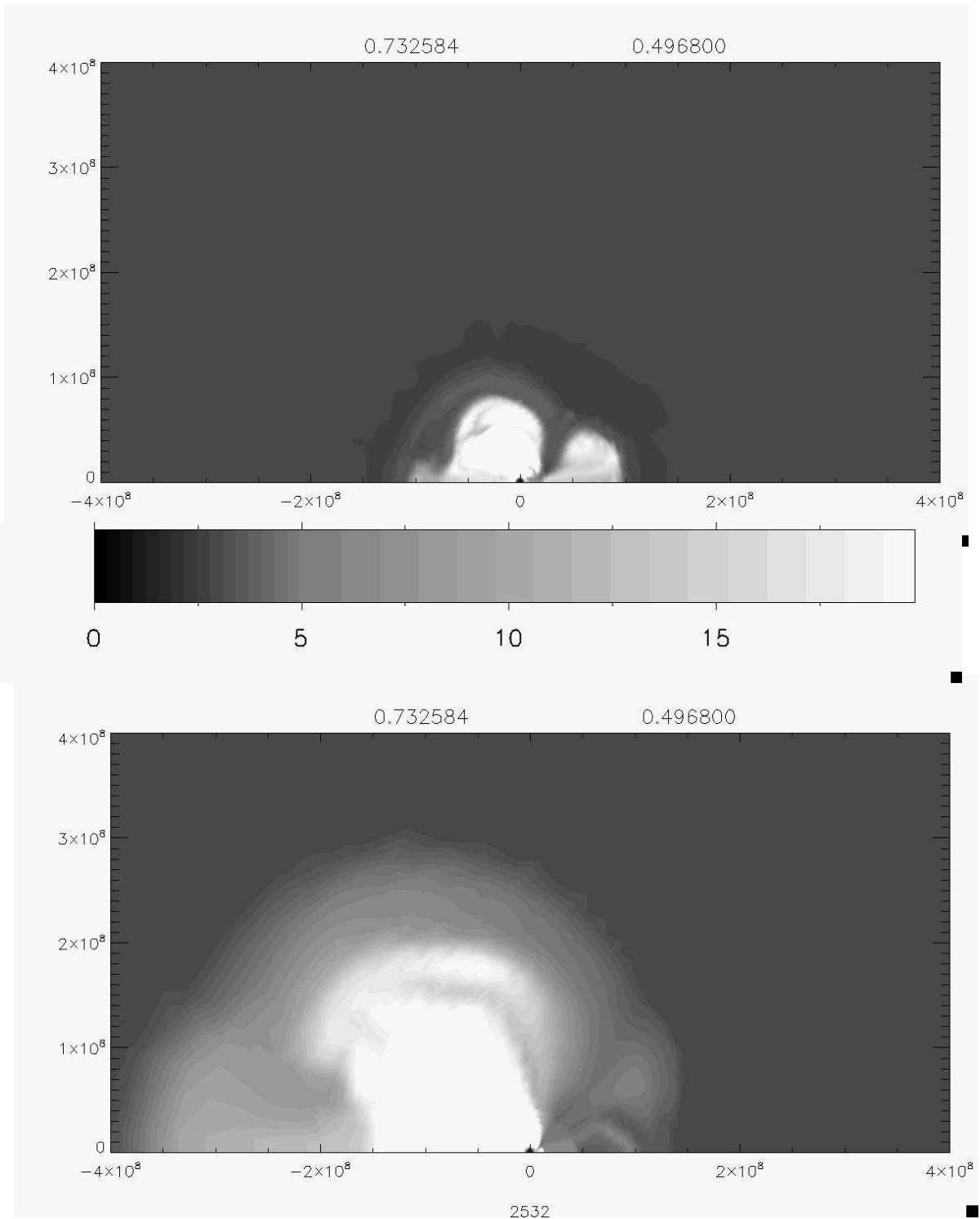}
\caption{In this figure, the state of the shock is displayed for the 11 M$_\odot$ model at $\sim$ 500 ms after bounce for the models with the nuclear network off (top) and on (bottom). In the former case, an explosion is not obtained, whereas in the latter, explosion does in fact occur. The variable shown is entropy. The x and y axes mark distance in cm. The disparate shock radii at this snapshot after bounce are evident. With the network on, the shock has reached $\sim$ 4000 km.
}
\end{figure}

\begin{figure}
\includegraphics[height=.175\textheight]{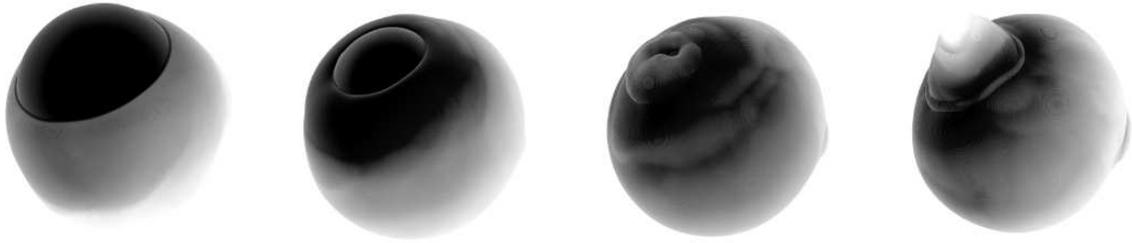}
\caption{The transition from the linear to the nonlinear phase in the development of the SASI in three dimensions illustrates the fundamental breakdown of axisymmetry, which in turn leads to the one-sided internal shock and the subsequent development of two strong counter rotating flows between the shock and the proto-neutron star, the innermost of which is responsible for spinning up the proto-neutron star. In the above figure, the Mach reflection is seen converging on the polar axis, but the convergence is not perfectly axisymmetric, resulting in the subsequent asymmetry between the reflected flows with respect to the azimuth.}
\end{figure}

\begin{figure}
\includegraphics[height=.4\textheight]{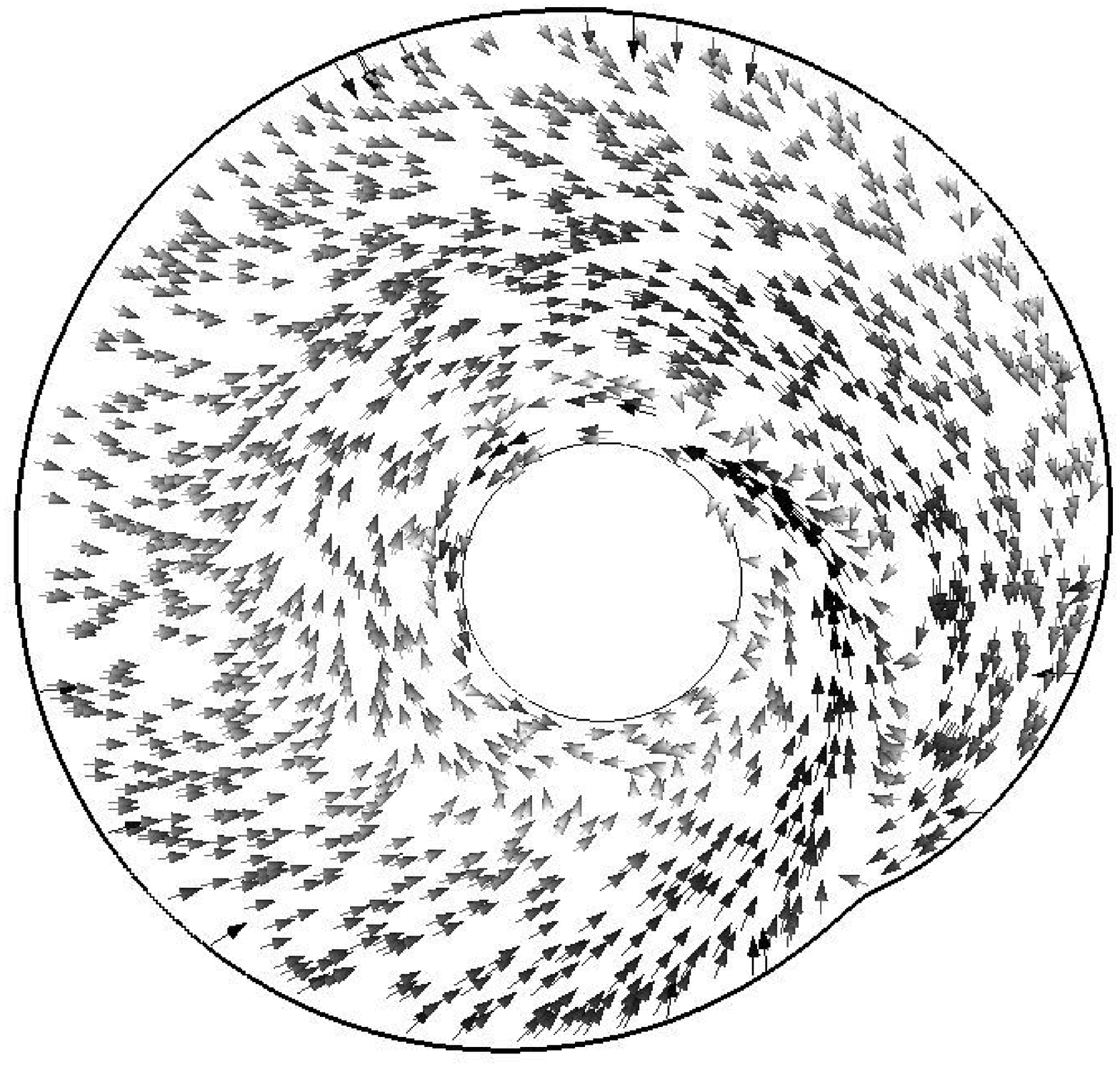}
\caption{The two counter rotating flows resulting from the non-axisymmetry of the Mach reflection and internal shock are evident here. The internal shock, rotating in a clockwise fashion, leads to a clockwise rotating flow behind it. The shock, however, subverts the material flowing through the accretion shock ahead of it into a counterclockwise rotating flow. This inner flow will in turn spin up the proto-neutron star, and this mechanism may be responsible for the observed spin periods of pulsars \cite{bm05Nature}.}
\end{figure}

\section{Conclusions}

The results we present here clearly indicate that we should not expect to pin down the core collapse supernova mechanism without three-dimensional models that include all of the presently known relevant physics. Our two-dimensional models yielded the surprise that nuclear burning qualitatively changes the outcomes despite its somewhat subtle influence on the shock. In our models, the confluence of neutrino heating, the SASI, and nuclear burning produced explosions, at least in the Newtonian case. The coupling of these individual components in this nonlinear problem had an {\it a priori} unpredictable outcome. To further drive home the point, the results of our three-dimensional simulations demonstrate that supernova dynamics will be fundamentally different in two and three spatial dimensions. The lack of an artificially imposed symmetry in three dimensions admits degrees of freedom not present in axisymmetric systems. These not only affect the shock dynamics and explosion but lead to new predictions with regard to supernova-associated phenomena such as neutron star spin and pulsar birth. Thus, while the consistent explosion of our two-dimensional Newtonian models with nuclear burning is an advance (no such consistency has been reported before in the literature), our three-dimensional models and the knowledge that the gravitational fields will deviate significantly from Newtonian, as well as the knowledge that we have in the models reported here completely ignored the stellar core magnetic fields, temper our conclusions and tell us that three-dimensional extensions of our two-dimensional models will be needed before any final conclusions can be made with regard to the role burning will play in the supernova mechanism. Nonetheless, our findings warn us that nuclear burning, modeled with sufficient realism, should no longer be ignored in future supernova models. Our results also tell us that future simulations will need to be carried out well beyond bounce ($t > 500$ ms) to determine final outcomes and that a revised paradigm for shock revival may be required: the shock may be revived after it exits the iron core, not while it is in the iron core.

\begin{theacknowledgments}
This work was supported by a SciDAC grant from the US Department of Energy High Energy, Nuclear Physics, and Advanced Scientific Computing Research Programs. A.M. is supported at the Oak Ridge National Laboratory, managed by UT-Battelle, LLC, for the US Department of Energy. The simulations presented here were performed at the Leadership Computing Facility at ORNL. We thank the National Center for Computational Sciences at ORNL for their resources and support. 
\end{theacknowledgments}




\end{document}